\documentclass{emulateapj}

\newcommand{\be}{\begin{eqnarray}}
\newcommand{\ee}{\end{eqnarray}}
\newcommand{\beq}{\begin{equation}}
\newcommand{\eeq}{\end{equation}}
\def\simless{\mathbin{\lower 3pt\hbox
      {$\rlap{\raise 5pt\hbox{$\char'074$}}\mathchar"7218$}}}
\def\simgreat{\mathbin{\lower 3pt\hbox
      {$\rlap{\raise 5pt\hbox{$\char'076$}}\mathchar"7218$}}} 

\renewcommand{\vec}[1]{\mbox{\boldmath $\displaystyle #1$}}

\newcommand{\grad}{{\mbox{\boldmath $\nabla$}}}

\begin{document}

\title{ Thermal Tides in Fluid Extrasolar Planets }

\author{Phil Arras\altaffilmark{1} and Aristotle Socrates\altaffilmark{2} }

\altaffiltext{1}{Department of Astronomy, University of Virginia,
P.O. Box 400325, Charlottesville, VA 22904-4325}

\altaffiltext{2}{Institute for Advanced Study,
Einstein Drive, Princeton, NJ 08540}

\email{ socrates@ias.edu, arras@virginia.edu }


\begin{abstract}

Asynchronous rotation and orbital eccentricity lead to time-dependent
irradiation of the close-in gas giant exoplanets -- the hot Jupiters.
This time-dependent surface heating gives rise to fluid motions which
propagate throughout the planet. We investigate the ability of this
``thermal tide" to produce a quadrupole moment which can couple to the
stellar gravitational tidal force. While previous investigations
discussed planets with solid surfaces, here we focus on entirely fluid
planets in order to understand gas giants with small cores. The Coriolis 
force, thermal diffusion and self-gravity of the perturbations
are ignored for simplicity. First, we
examine the response to thermal forcing through analytic solutions of
the fluid equations which treat the forcing frequency as a small
parameter. In the ``equilibrium tide" limit of zero frequency, fluid
motion is present but does not induce a quadrupole moment. In the next
approximation, finite frequency corrections to the equilibrium tide do
lead to a nonzero quadrupole moment, the sign of which torques the
planet {\it away} from synchronous spin. We then numerically solve the
boundary value problem for the thermally forced, linear response of a
planet with neutrally stratified interior and stably stratified
envelope. The numerical results find quadrupole moments in agreement with the
analytic non-resonant result at sufficiently long forcing
period. Surprisingly, in the range of forcing periods of 1-30 days, the
induced quadrupole moments can be far larger than the analytic result
due to response of internal gravity waves which propagate in the
radiative envelope. We discuss the relevance of our results for the
spin, eccentricity and thermal evolution of hot Jupiters.

\end{abstract}



\section{Introduction}

Many of the gas giant exoplanets orbiting close to their parent stars --
the hot Jupiters -- are observed to have radii far larger than the radius
of Jupiter, $R_J$, implying high temperatures deep in the planetary
interior \citep{2000ApJ...534L..97B}.  A powerful internal heat source
must be present in order to prevent their rapid contraction. Tidal heating
is a possible solution.  However, the gravitational tide synchronization
and circularization timescales are much shorter than the age of the 
system if the tidal quality factor is comparable to that of Jupiter. Tidal 
heating could then power the observed radii only for a brief period, 
far shorter than the age of observed systems.

We propose that time-dependent thermal forcing of hot Jupiters leads 
to a ``thermal tide" torque pushing the planet away from synchronous 
rotation, and possibly circular orbits. The equilibrium spin state
is set by the competition between the opposing thermal and gravitational
tide torques.  Gravitational tide dissipation continues to operate 
in this torque equilibrium, implying a {\it steady state} heat source to power
the large radii. The thermal tide torque mechanism was initially proposed 
by Gold and Soter (1969, GS from here on) in order to explain Venus' slow, 
retrograde spin.  Here, we generalize their analysis for fluid planets in which 
the central solid core is either non-existent or dynamically unimportant.
 
The existence of dynamically important thermal tide torques in fluid planets 
has recently been cast in doubt (Goodman 2009 and  Gu \& Ogilvie 2009). These
authors argue that at forcing frequencies smaller than the planet's dynamical
frequency, the quadrupole moment will be negligible due to 
isostatic adjustment. Furthermore, the former author asserts that the torque
will push the planet toward synchronous rotation, rather than away. A careful
study of the fluid motion induced by surface thermal forcing in fluid gas giant
planets is needed to address the following questions: 
What is the direction and depth dependence of the resulting flow?
What is the relevance of the 
concept of ``isostatic adjustment" to this inherently time-dependent problem? 
What sets the ``dynamical frequency below which the quadrupole moments 
become small? What is the sense of the torque (synchronous versus 
asynchronous)? What is the magnitude of the torque? Once these questions have
been answered, the ability of thermal tide torques to promote asynchronous rotation
and tidally inflated radii can then be assessed.

Here, we study the fluid flow and resulting quadrupole moments
induced by thermal forcing at the surface of a gas giant planet. To
isolate what we believe to be the relevant physical mechanisms, we
ignore thermal diffusion, the Coriolis force and self-gravity of the
perturbations, while utilizing a simplified model for the planetary
structure.

The plan of this paper is as follows.  In \S\ref{s:motivation} we provide
motivation for performing the ensuing calculations. We then describe the problem
set-up and details of the background model for the fluid planet in \S
\ref{s:setup}. \S \ref{sec:horiz} reviews the relation between density
perturbation and quadrupole moment, and the torque on the planet and
orbit in terms of the quadrupole moment. The linearized fluid equations
are presented in \S \ref{s:boundary}.   In \S \ref{ss: zero_freq} and
\ref{ss: finite} we develop analytic solutions in the zero and small,
but finite, frequency limit.  Numerical results to the linearized
boundary value problem are presented in \S\ref{sec:numerical}. In
\S\ref{s: resonant}, the numerical results are explained in terms of
the response of envelope gravity waves that are excited thermal forcing.
We summarize in \S\ref{s: summary}.


\section{Motivation }
\label{s:motivation}

\begin{figure}
\epsscale{1.0}
\plotone{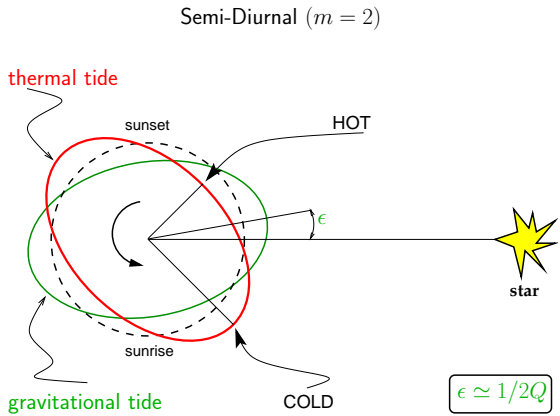}
\caption{  Geometry of the semi-diurnal ($m=2$)
thermal and gravitational tidal density perturbations.
The planet rotates counterclockwise.
In order for the torque to
push the planet away from synchronous spin, the density
perturbation must {\it lead} the line joining planet and star.}
\label{fig:cartoon}
\end{figure}

We begin with a brief discussion of the possible role of thermal tides
for the rotation rates and radii of hot Jupiters.  

Thermal forcing may induce a ``bulge" which is misaligned with star-planet
line (see figure \ref{fig:cartoon}). The misalignment is a result
of thermal inertia, the tendency of maximum temperature to lag maximum
heating. Since insolation is finite on the day side and zero on the night
side, thermal forcing is spread over a range of harmonics. In particular,
the semi-diurnal ($m=2$) component of the thermal forcing may induce
a quadrupole moment in the fluid, which can then couple to stellar
gravitational field. Therefore, this ``thermal tidal torque'' can, in principle,
push the planet away from a synchronous spin state. The equilibrium spin
rate is then set by a balance of the thermal tide torque, promoting
asynchronous spin, and the gravitational tide torque, which acts to
push the spin back toward the synchronous state.  The two bulges are
illustrated in figure \ref{fig:cartoon}.  

In the asynchronous spin state set by torque equilibrium, the
gravitational tide generates heat in a steady state manner. If the
dissipation occurs sufficiently deep within the convective core of the
planet, a thermal equilibrium will be established between (gravitational)
tide heating of the core and heat loss at the radiative-convective
boundary. To explain the large observed radii, the equilibrium core
entropy must be much larger than that of a passively cooling planet at
the same age.

The thermal tide torque is expected to deposit angular momentum mainly
in the surface radiative layer. Were there no sink of angular momentum,
the total angular momentum of the planet would increase indefinitely.
If we again assume that gravitational tide dissipation occurs deep
in the convective core, then angular momentum must be transported
between the thermal tide source region and the gravitational tide sink
region. Presumably this angular momentum transport would be accomplished
through (turbulent) viscous forces between neighboring fluid layers.

We estimate the importance of thermal tides in hot Jupiters by comparing
the quadrupole moment induced by the thermal tide to that from dissipation
acting on the gravitational tide. For the purposes of a fiducial
estimate, we use the thermal tide quadrupole formula relevant for Venus
by \citet{1969Icar...11..356G}:
\be Q^{\rm (th)}\sim \frac{\Delta MR^2_p}{\sigma t_{\rm th}}.
\label{eq:GSapprox}
\ee 
Here $\Delta M$ is the mass down to the photospheric layer for the
starlight, $R_p$ is the planet's radius, $\sigma$ is the tidal forcing
frequency and the thermal time of the absorbing layer is given by
\be 
t_{\rm th}\sim \frac{\Delta M c_pT}{R^2_pF_*},  
\ee
where $c_p$ is the specific heat at constant pressure, $T$ is the
temperature, and $F_\star$ is the stellar flux.  The frequency dependence
in eq.\ref{eq:GSapprox} reflects that the temperature and density
perturbations in the atmosphere become small when the forcing period
is short.  The accuracy with which eq.\ref{eq:GSapprox} represents a
fluid atmosphere is the subject of the remainder of the paper.

The quadrupole moment induced by the 
stellar gravitational tide which leads to secular evolution is given by 
(e.g. \citealt{1966Icar....5..375G})
\be
Q^{\rm (grav)}\sim\left(\frac{h_t}{R_p}\right)\left(\frac{\sigma}{n}
\right)\frac{M_pR^2_p}{{\cal Q}_{p}}
\ee
where the height of the tidal bulge is $h_t \sim R_p (M_\star/M_p)(R_p/a)^3$,
the orbital frequency is $n \simeq (GM_\star/a^3)^{1/2}$, the semi-major axis
is $a$, the stellar mass is $M_\star$, and the tidal quality factor of the 
planet is ${\cal Q}_p$.
Here $\sigma\propto n-\Omega$ is proportional to the departure of the rotation rate
$\Omega$ from synchronous rotation, $\Omega=n$.
As a fiducial value, 
the tidal quality factor of Jupiter is constrained to be ${\cal Q}_p\sim 10^5-10^6$ (Goldreich \& Soter 1965;
Yoder \& Peale 1981).  

The ratio between the thermal tide and gravitational tide 
quadrupole is given by 
\be
\frac{{Q}^{\rm (th)}}{{Q}^{\rm (grav)}}
& \sim  & \left( \frac{n}{\sigma} \right)^2 
\left( \frac{\Delta M}{M_p} \right)
\left( \frac{M_p}{M_\star} \frac{a^3}{R_p^3}\right)
\left( \frac{ {\cal Q}_p }{n t_{\rm th}} \right)
\nonumber \\ & \simeq & 
\left( \frac{n}{\sigma} \right)^2
\left( \frac{\Delta M}{10^{-8} M_p} \right)
\left( \frac{M_p}{10^{-3}M_\star} \right)\
\nonumber \\ & \times & \left( \frac{a}{10^2 R_p} \right)^3
\left( \frac{ {\cal Q}_p }{10^5} \right)
\left( \frac{1}{n t_{\rm th}} \right).
\label{eq:quad_compar}
\ee
The fiducial estimate in eq.\ref{eq:quad_compar} shows that thermal 
and gravitational tidal effects are competitive for hot Jupiters, 
with large departures from synchronous spin $\sigma \sim n$
expected.

Given the equilibrium spin rate, the gravitational tide will generate
heat at a rate
\be
\dot{E}^{(\rm GT)} & = & \frac{\sigma}{m}n^2Q^{(\rm grav)}=
-\frac{\sigma}{m} n^2Q^{(\rm th)}\\
& \simeq & 10^{28}\,{\rm erg}\,\,{\rm s}^{-1}\nonumber\left(
\frac{\sigma}{n}\right)\left(\frac{{\rm 3\, days}}{P_{\rm orb}}\right)^3
\left(\frac{Q^{\rm (th)}}{10^{-8}M_JR^2_J}\right).
\label{e: dissipation_rate}
\ee
This fiducial estimate of the heating rate is comparable, or larger than, the
core cooling rates found in \citet{2006ApJ...650..394A} for many ``problem"
planets, which cannot be explained by passive cooling. If this tidal heating
is deposited sufficiently deep in the core, it may then prevent these planets from
contracting, explaining the inflated radii of some hot Jupiters.

\section{ Problem set up}\label{s:setup}

In this section we expand the thermal and gravitational forcing terms
in terms of a Fourier expansion, define the co-ordinate system used,
and discuss the background planetary structure model.

\subsection{description of time-dependent thermal and gravitational 
forcing}

We seek solutions to the boundary value problem describing linear
response of a gaseous planet to a gravitational tidal acceleration $-\grad
U(\vec{x},t)$ and an imposed entropy perturbation, $\Delta s(\vec{x},t)$,
that results from time-dependent insolation. 

Consider a planet of mass $M_p$ and radius $R_p$ in a circular orbit
of separation $a$ around a solar-type star of mass $M_\star=M_\odot$,
radius $R_\star=R_\odot$ and effective temperature $T_\star=5800\ {\rm
K}$. The bolometric flux at the planet's surface is then $F_\star=\sigma_{\rm sb}
T_\star^4(R_\star/a)^2$, where $\sigma_{\rm sb}$ is the Stefan-Boltzmann constant.
The orbital frequency is $n=(GM_\star/a^3)^{1/2}$ and the  planet's spin
frequency is $\Omega$. The orbital phase in the frame corotating with the planet
will be denoted $\Phi=(n-\Omega)t$. 

It is convenient to express $U(\vec{x},t)$ and $\Delta s(\vec{x},t)$ 
in a Fourier series in time and spherical harmonics in 
angle i.e., 
\be
\Delta s(\vec{x},t)=\sum_{\ell m}\Delta s_{\ell m}(r)
Y_{\ell m}(\theta,\phi) e^{-i \sigma_{m} t}
\ee  
and 
\be
U(\vec{x},t)=\sum_{\ell m}U_{\ell m}(r)Y_{\ell m}
(\theta,\phi) e^{-i \sigma_{m} t}.
\label{eq:Uexp}
\ee
The forcing frequency is then
$\sigma_{m}=m(n-\Omega)$.
\footnote{ For an eccentric orbit, this expression
is generalized to $\sigma_{mk}=kn-m\Omega$ where $k$ is an integer.} For
$\ell=2$, only $m=\pm 2$
are allowed due to the even parity of $U$ and $\Delta s$.  
We note that while the longitude-dependent stellar irradiation is zero on the
night side and nonzero on the day side, the individual Fourier components
are nonzero on both day and night sides. By adding together many Fourier 
components one recovers the true position-dependent flux.

Lagrangian and Eulerian variations are denoted by $\Delta$ and $\delta$,
respectively. The Lagrangian entropy perturbation, $\Delta s$, is a
convenient variable to use as it varies solely due to non-adiabatic
processes, here the time-dependent heating due to insolation. By contrast,
the Eulerian entropy perturbations can vary due to motion parallel
to a background entropy gradient, and the temperature can vary due to
pressure perturbations.

Unfortunately, the governing equations are not separable when the
Coriolis force is included. Due to this complication, for simplicity,
we neglect the Coriolis force. By doing so, the response for each
individual spherical harmonic can studied separately. We will suppress
the subscripts $(\ell,m)$ where no confusion will arise.

Scalar quantities are represented as 
$f(\vec{x},t)=f(r)Y_{\ell m}(\theta,\phi)e^{-i \sigma t}$,
and the Lagrangian displacement vector, $\vec{\xi}(\vec{x},t)$ is 
expanded in the poloidal
harmonics
\be
\vec{\xi}(\vec{x},t)=\left[\xi_{r}(r)\,\vec{e}_r
+\xi_{h}(r)r\grad\right]Y_{\ell m}(\theta,\phi) e^{-i \sigma t}.
\ee

The entropy perturbation $\Delta s$ is related to the Lagrangian density and 
pressure perturbations, $\Delta \rho$ and $\Delta p$ respectively, 
via the thermodynamic relation
\be
\frac{\Delta \rho}{\rho} & = & \frac{\Delta p}{\Gamma_1 p} + 
\rho_s \frac{\Delta s}{c_p}.
\label{eq:eos_lagr}
\ee
Here $\Gamma_1=(\partial \ln P/\partial \ln\rho)_s$ is the first adiabatic
index, $c_p=T (\partial s/\partial T)_p$ is the specific heat per 
unit mass at constant
pressure, and $\rho_s=c_p (\partial\ln \rho/\partial s)_p$ 
is the negative of the 
volume expansion coefficient at constant pressure. For an ideal 
gas, $\rho_s=-1$.

If the thermal inertia of the absorbing layer is large, then 
thermal diffusion can be ignored on the timescale $\sigma^{-1}$
of the perturbation. 
From here on, we
assume that this limit applies.  Perturbations of the 
heat equation then take on the simple form
\be
\frac{\Delta s}{c_p}=\frac{i}{\sigma}\frac{\delta\epsilon}{c_pT}.
\label{e: first}
\ee
Here $\delta \epsilon$ is the perturbed rate of heating, per unit
mass. The depth dependence of
$\delta\epsilon$ is approximated as
\be
\delta\epsilon(p) & = & \kappa_\star F_\star
\exp(-\kappa_\star p/g)
\label{eq:epsilon}
\ee
for the $\ell=|m|=2$ harmonics of interest.  Here, $\kappa_\star$ is the
effective opacity for where starlight is absorbed, at a characteristic
pressure $g/\kappa_\star$.  The thermal time at the base of the
heated layer is $t_{\rm th}=c_pT/\kappa_\star F_\star$. 
Therefore,
the imposed entropy perturbation can be written as 
\be 
\frac{\Delta s}{c_p} &
= & \frac{i}{\sigma t_{\rm th}} \exp(-\kappa_\star p/g).
\label{eq:Deltas}
\ee
For semi-diurnal forcing of an
asynchronously rotating planet, the stellar insolation $\delta\epsilon$
achieves its maximum value at noon and at midnight.
The entropy perturbation $\Delta s$ then attains
maxima at 3PM and 3AM, and minima at 9AM and 9PM.

The $\ell=2$ coefficients for the tidal potential are given
by 
\be
U_{2m}=-\sqrt{ \frac{3\pi}{10} } n^2 r^2
\ee
for both $m=\pm 2$.

\subsection{Background Model}

\begin{figure}
\epsscale{1.2}
\plotone{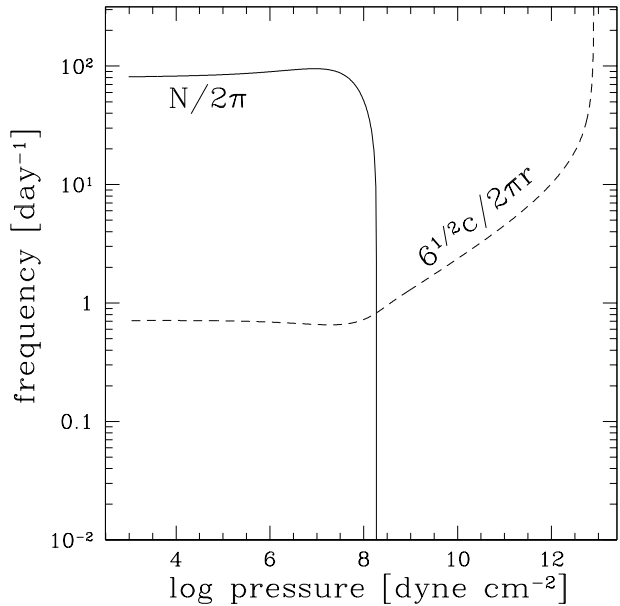}
\caption{ Propagation diagram showing Brunt-Vaisalla frequency
($N$, solid line) and Lamb frequency for $\ell=2$ ($6^{1/2}c/r$, dashed line)
as a function of  depth. An acoustic wave
with a fixed frequency $\sigma$ can propagate in regions where
$\sigma > N,ck_\perp$ and gravity waves can propagate where
$\sigma < N,ck_\perp$. Here $c$ is the adiabatic sound speed and
$k_\perp=\sqrt{6}/r$ is the horizontal wavenumber. The region near the surface with $N^2>0$ is
stably stratified, and the central region with $N^2 \simeq 0$
is neutrally stratified. }
\label{fig:prop}
\end{figure}

Hot Jupiters possess a convective core, blanketed by a thin
radiative
envelope at the surface(e.g. \citealt{2000ApJ...534L..97B},
\citealt{2006ApJ...650..394A}). Rather than construct a model with a
detailed equation of state and radiative transfer, we use a simple
parametrized equation of state which enforces neutral stratification 
in the core and stable stratification with roughly constant scale height
in the envelope. The equation of hydrostatic balance is integrated using
\be
\rho(p) & =&
e^{-p/p_b} \left( \frac{p}{a^2} \right) + \left(1 - e^{-p/p_b} 
\right) \sqrt{\frac{p}{K_c}}.
\label{eq:eos}
\ee
The constants $p_b$, $K_c$ and $a=(p_bK_c)^{1/4}$ are the pressure at the 
base of the radiative
envelope, parameter setting the entropy of the core and the isothermal sound 
speed of the envelope, respectively. The factors with $e^{-p/p_b}$ enforce
a narrow transition at the radiative-convective boundary. The core 
compressibility is set to $K_c=GR_J^2$.
The propagation diagram for a model with $p_b=100$ bar is shown in figure \ref{fig:prop}. The 
mass and radius are $M_p=0.7M_J$ and $R_p=1.27R_J$.
The mass of the radiative envelope is roughly $p_b/p(r=0) = 10^{-5}$ 
of the planet's total mass.  Though this model is simple, 
it accurately reproduces the gross features of more detailed 
models (cf. Arras \& Bildsten).

To compute the sound speed we take $\Gamma_1=2$ so that the 
core is neutrally stratified.
The Brunt-Vaisalla frequency is then
\be
N^2 & =& -g\left( \frac{d\ln \rho}{dr} - \frac{1}{\Gamma_1} 
\frac{d\ln p}{dr} \right) 
\nonumber \\
& = & \frac{g}{H} \left( \frac{d\ln \rho}{d\ln p} - \frac{1}{\Gamma_1}  \right)
\label{eq:nsqmodel}
\ee
where $g(r)=Gm(r)/r^2$ is the gravity, $H=p/\rho g$ is the pressure
scale height, and the derivative $d\ln \rho/d\ln p$ is computed
analytically from eq. (\ref{eq:eos}). 


\section{Quadrupole moment and tidal torques}
\label{sec:horiz}

The interaction Hamiltonian, which coupled the density field of the planet to
the gravitational tidal potential of the star is given by
(e.g. \citealt{newcomb62})
\be
H & = & \int d^3x^* \rho^*(\vec{x}^*,t) U(\vec{x}^*,t).
\label{eq:H}
\ee
Here $\vec{x}^*$ and $\rho^*$ are the position and mass density of 
a fluid element in the perturbed planet.
The fluid position in the background planet, $\vec{x}$, and in the 
perturbed planet, $\vec{x}^*$,
are related by $\vec{x}^*=\vec{x} + \vec{\xi}(\vec{x},t)$, where 
$\vec{\xi}$ is the Lagrangian
displacement vector. Mass conservation implies 
$d^3x^* \rho^*(\vec{x}^*,t)=d^3x\rho(\vec{x})$, where the 
volume integral is performed over the background planet. 
Eq.(\ref{eq:H}) may then be expanded as
\be
H & = & \int d^3x \rho(\vec{x}) U(\vec{x}+\vec{\xi},t)
\nonumber \\ & \simeq & 
\int d^3x \rho(\vec{x}) \left[ U(\vec{x}) +  \vec{\xi} \cdot \grad U(\vec{x},t)
+ {\cal O}(\xi^2) \right].
\label{eq:H2}
\ee
The first term in eq. (\ref{eq:H2}) corresponds to the interaction energy
of the background, and can be ignored since it is independent of $\xi$.
The second term is the interaction energy of the perturbations at linear order
and is the term of interest. We ignore nonlinear terms of order $\xi^2$ on the 
assumption that they are small.

To evaluate eq. (\ref{eq:H2}), we expand the tidal potential in spherical
harmonics (see eq.\ref{eq:Uexp})
\be
U(\vec{x},t) & = & - GM_\star \sum_{\ell m} \frac{4\pi}{2\ell +1}
\left( \frac{r^\ell}{D^{\ell+1}} \right)
Y_{\ell m}(\pi/2,\Phi) Y_{\ell m}^*(\theta,\phi)
\label{eq:U}
\ee
and define the time-dependent multipole moments of the planet
\be
Q_{\ell m}(t) & =& \int d^3x\ \rho\vec{\xi}(\vec{x},t) 
\cdot \grad \left( r^{\ell} Y^*_{\ell m}(\theta,\phi) \right).
\label{eq:Qlmdef}
\ee
Here $D$ and $\Phi$ are the orbital separation and phase.
Integrating by parts and using the continuity equation for the
perturbations
\be
\delta \rho & = & - \grad \cdot \left( \rho \vec{\xi} \right),
\label{eq:continuity}
\ee
eq. (\ref{eq:Qlmdef}) becomes
\be
Q_{\ell m}(t) & =& \int d^3x\ r^{\ell} Y^*_{\ell m}(\theta,\phi) 
\delta \rho(\vec{x},t).
\label{eq:Qlm2}
\ee
Hence it is the Eulerian density perturbation which is needed for the
quadrupole moment.

Since $r^\ell \delta\rho(\vec{x},t)$ is a real quantity, and
$Y_{\ell m}^*=(-1)^m Y_{\ell, -m}$, the moments must satisfy
$Q_{\ell m}^* = Q_{\ell, -m} (-1)^m$.  By defining $W_{\ell
m}\equiv[4\pi/(2\ell+1)]Y_{\ell m}(\pi/2,0)$, the interaction Hamiltonian
in eq. (\ref{eq:H}) may be conveniently expressed in terms of a sum over
spherical harmonics as
\be
H(D,\Phi) & = & -GM_\star \sum_{\ell m} W_{\ell m}
Q^*_{\ell m}(t) \frac{e^{-im\Phi}}{D^{\ell+1}}.
\ee

The rate of change of orbital angular 
momentum, ${L}_{\rm orb}=GM_pM_\star(Ga/(M_p+M_\star))^{1/2}$, 
and the torque on the planet, ${\mathcal N}$, are then
\be
\dot{L}_{\rm orb} & = & - {\mathcal N} = 
- \frac{\partial}{\partial \Phi} H(D,\Phi).
\label{eq:Lorbdot}
\ee
The secular evolution of the planet-orbit system is due to the
presence of quadrupole moments that are out of
phase with the tidal acceleration i.e., the imaginary component of
the quadrupole.
The torque on the planet for a circular orbit is dominated by the
semi-diurnal term with $|m|=2$ and forcing frequency $2(n-\Omega)$. The torque
at this order is
\be
\mathcal{N} & = & 4 \left( \frac{3\pi}{10} \right)^{1/2} 
\left( \frac{M_p+M_\star}{M_\star} \right) n^2
{\rm Im} \left( Q_{22} \right)
\nonumber \\ & \simeq &
4n^2 {\rm Im} \left( Q_{22} \right).
\label{eq:Ntt2}
\ee
The sign in eq. (\ref{eq:Ntt2}) is such that density perturbations
which {\it lead} maximum heating tend to torque the planet {\it away} from
synchronous rotation, and vice versa.

\section{Solution of the boundary value problem}
\label{s:boundary}

Conservation of mass and momentum, the first law of thermodynamics,
and two boundary conditions are employed
in order to obtain the planet's response to thermal and gravitational 
forcing. First we  develop equations that can be solved as a 
boundary value problem. We then 
show that planet's response can be represented
as a sum of eigenmodes. The governing equations
below are standard and can be found in, for example, 
\citet{1989nos..book.....U}.
 
Conservation of horizontal and vertical momentum are given by 
\be
- \sigma^2 \xi_h & = & - \left( \frac{\delta p/\rho + U}{r} \right)
\label{e: mom_h}
\ee
and
\be
- \sigma^2 \xi_r & = & - \frac{1}{\rho}\frac{d\delta p}{dr} - 
g \frac{\delta \rho}{\rho}
- \frac{dU}{dr},
\label{e: mom_r}
\ee
respectively.
Here $\delta p$ and $\delta \rho$ are the Eulerian pressure and density
perturbations.
We ignore the self-gravity of the perturbations for simplicity.
The continuity eq. (\ref{eq:continuity}) may be written as
\be
\frac{\delta \rho}{\rho} & = & -  \frac{1}{r^2\rho} \frac{d}{dr}
\left( r^2 \rho \xi_r \right)
+ \frac{\ell(\ell+1)}{r} \xi_h
\label{eq:continuity1}
\\ & = & 
- \frac{1}{r^2\rho} \frac{d}{dr}
\left( r^2 \rho \xi_r \right)
+ \frac{\ell(\ell+1)}{\sigma^2 r^2} \left( \frac{\delta p}{\rho}
+ U \right).
\label{eq:continuity2}
\ee
where we have used eq. (\ref{e: mom_h}).
By substituting the relation between Lagrangian and Eulerian quantities i.e.,
$\Delta = \delta  + \vec{\xi}\cdot\grad $, eq. (\ref{eq:eos_lagr})
becomes
\be
\frac{\delta \rho}{\rho} & = & \frac{\delta p}{\rho c^2} + 
\frac{N^2}{g} \xi_r + \rho_s \frac{\Delta s}{C_p}
\label{e: EOS}
\ee
where $c^2=\Gamma_1p/\rho$ is the adiabatic sound speed.
The imposed entropy perturbations drive vertical fluid motion
through buoyancy forces $-g\rho \rho_s \Delta s/c_p$.
The vertical fluid motion in turn drives horizontal motion.

We solve the system of 3 equations (\ref{e: mom_r}),
(\ref{eq:continuity2}) and (\ref{e: EOS}), for the three variables
$\delta p$, $\delta \rho$ and $\xi_r$. While $\delta\rho$ can 
be eliminated as well, we find that better numerical
precision is achieved by keeping $\delta \rho$ in the system of equations.

Finally, we remark on the boundary conditions. 
Eq.(\ref{e: EOS}) is algebraic, so no
boundary condition is needed as this equation can be 
evaluated on the boundaries.
At the center of the planet, we require all variables to be finite.
This implies (e.g. \citealt{1989nos..book.....U};
take $\xi_r \propto r^{\ell-1}$ and $\delta p,\delta \rho \propto r^\ell$
in eq.\ref{e: mom_r})
\be
\sigma^2 \xi_r & = &  \frac{\ell}{r} \left( \frac{\delta p}{\rho} + U \right).
\label{eq:bc1}
\ee
We present results for two different upper boundary conditions.
The ``standard" boundary condition requires that the Lagrangian pressure
perturbation $\Delta p$ vanishes \citep{1989nos..book.....U}
\be
\delta p/\rho & = & g \xi_r.
\label{eq:bc2}
\ee
If the fluid perturbation is evanescent, this boundary 
condition is valid. 
We also present results for an ``outgoing wave" boundary 
condition, in which wave energy propagates outward at the upper
boundary (e.g. \citealt{1989nos..book.....U}).
We implement this outgoing wave boundary condition as follows. The
atmosphere above the upper boundary is idealized as isothermal, with
constant $H$, $N$, $g$, $c$ and horizontal wavenumber
$k_\perp=\sqrt{\ell(\ell+1)}/R_p$.  In an isothermal atmosphere, the
fluid variables $\delta p/\rho,\xi_r \propto e^{bz}$
(cf. \citealt{1990ApJ...363..694G}), where $z$ is altitude and the
complex constant
\be
b & = & \frac{1}{2H} \pm i \left[ k_\perp^2 \frac{N^2}{\sigma^2} - k_\perp^2 + \frac{\sigma^2}{c^2}
- \frac{1}{4H^2} \right]^{1/2}
\label{eq:s}
\ee
determines the run with height. When the argument of the square root is positive,
the wave is propagating. The sign of the second order wave flux is
\be
F_{\rm wave}= \delta p\,\dot{\xi_r} & \propto & 
- \left( \frac{\sigma}{N^2-\sigma^2}\right) {\rm Im}(b).
\ee
We choose outgoing wave flux by choosing the appropriate sign of ${\rm Im}(s)$ 
to make
$F_{\rm wave} >0$, given the signs of $\sigma$ and $N^2-\sigma^2$.
For the propagating case, the boundary condition is enforced as
\be
\frac{d}{dr} \left( \frac{\delta p}{\rho} \right) & = & b \left( \frac{\delta p}{\rho} \right).
\ee
When the argument of the square root in eq. (\ref{eq:s}) is negative,
the wave is evanescent and we enforce the ``hydrostatic" boundary
condition in eq. (\ref{eq:bc2}).

\subsection{Zero-frequency limit: the equilibrium tide}
\label{ss: zero_freq}

The equilibrium tide solution is found by setting $\sigma\rightarrow 0$ 
(e.g. \citealt{1989ApJ...342.1079G}).
The pressure perturbation is found from
eq. (\ref{e: mom_h}) to be
\be
\delta p^{\rm (eq)} & = & - \rho U.
\label{eq:deltapeq}
\ee
Substitution of eq. (\ref{eq:deltapeq})
into eq. (\ref{e: mom_r}) gives
\be
\frac{\delta \rho^{\rm (eq)}}{\rho} & = 
& \frac{d\ln \rho}{dr} \left( \frac{U}{g} \right).
\label{e: rho_eq}
\ee
Inserting eq.(\ref{eq:deltapeq}) and eq. (\ref{e: rho_eq})
into eq. (\ref{e: EOS}) gives
\be
N^2 \xi_r & = & - \frac{N^2}{g} U - g \rho_s \frac{\Delta s}{c_p}.
\label{eq:eqtideeos}
\ee
In stably stratified regions, $N^2>0$ and eq. (\ref{eq:eqtideeos})
can be solved for the vertical displacement
\be
\xi_r^{\rm (eq)} & = & - \frac{U}{g} - \frac{g}{N^2} \rho_s \frac{\Delta s}{C_p}.
\label{eq:xireq}
\ee
In the central convection zone, where $N^2 \simeq 0$, no solution exists
for eq. (\ref{eq:eqtideeos}), since the terms with $N^2$ go to zero while the 
$\Delta s$ term is nonzero.  It follows that the concept of an 
equilibrium tide breaks down in regions that are 
neutrally stratified and therefore, the forcing frequency cannot 
be safely set to zero.
That is, the fluid response in neutrally stratified regions 
is inherently a ``non-equilibrium tide."
Finally, substitution of eq.(\ref{eq:deltapeq}), (\ref{e: rho_eq}) 
and (\ref{eq:xireq}) into eq. (\ref{eq:continuity1})
yields the equilibrium horizontal displacement
\be
r \rho \xi_h^{\rm (eq)} & = & \frac{1}{\ell(\ell+1)}
\left[ \frac{d}{dr} \left( r^2 \rho \xi_r^{\rm (eq)} \right) 
+ r^2 \delta \rho^{\rm (eq)} \right].
\label{eq:xiheq}
\ee
Hence both vertical and horizontal motions exist in the equilibrium tide
limit, driven by gravity and entropy fluctuations. 

Clearly, the equilibrium tide density perturbation in eq. (\ref{e:
rho_eq}), and thus the equilibrium tide quadrupole moment, has
no contribution from time-dependent surface heating. Inspection of
eq. (\ref{e: EOS}) show that the terms $\propto\Delta s$ cancel out one
another.  The last term in eq. (\ref{e: EOS}) represents a density decrease
due to heating, which is precisely compensated by the $N^2\xi_r/g$
term that results from moving across surfaces of constant pressure.
In other words, denser fluid is brought up from below, exactly canceling
the local density decrease due to heating.

\begin{figure}
\epsscale{1.2}
\plotone{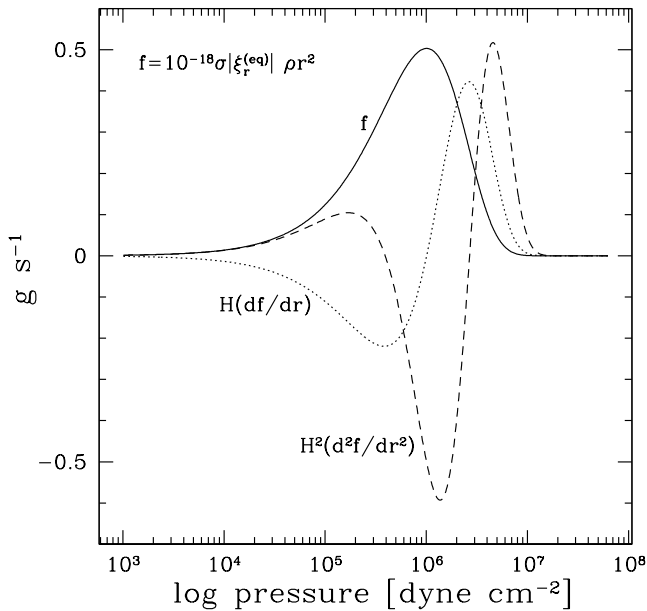}
\caption{ Plot of the function $f=10^{-18}\sigma |\xi^{\rm (eq)}_r| \rho r^2 = 10^{-18} \rho r^2
\frac{g}{N^2} \frac{\kappa_\star F_\star}{c_pT} e^{-\kappa_\star p/g}$
and its first and second derivative versus depth. Here $H=p/(\rho g)$ is the pressure
scale height. The first derivative is needed
for the horizontal displacement (eq.\ref{eq:xiheq2}).
In this plot, the base of the heating layer is at $g/\kappa_\star=1\ {\rm bar}$
and the base of the radiative zone is at $p_b=100\ {\rm bar}$.}
\label{fig:func}
\end{figure}

Time-dependent insolation heats the surface and initiates horizontal
motion. Consider subsynchronous rotation ($n-\Omega>0$) so that an observer at rest
with respect to the planet sees the star rotate with angle $\Phi(t)=(n-\Omega)t$
in the positive $\phi$
direction (see figure \ref{fig:cartoon}).
For $m=2$, the heating function $\delta \epsilon \propto \cos[2(\phi-\Phi)]$,
and maxima in the heating function occur at 
$\phi-\Phi=0,\pi$.
The entropy perturbation takes the form (eq.\ref{eq:Deltas})
\be
\Delta s(\vec{x},t) & = & \left( \frac{-1}{n-\Omega} \right) \left( \frac{\kappa_\star F_\star}{c_p T} e^{-\kappa_\star p/g} \right)
\nonumber \\ & \times & 
Y_{22}(\theta,0) \sin \left[ 2(\phi-\Phi) \right],
\ee
giving the hottest points ($\Delta s$ a maximum) at
$\phi-\Phi=-\pi/4, 3\pi/4$, and the coldest  points at $\phi-\Phi=\pi/4, -3\pi/4$.
This phase shift reflects the lag between maximum heating and maximum
temperature.  The displacement in the $\phi$ direction becomes
(eq.\ref{eq:xiheq})
\be
\xi_\phi(\vec{x},t) & = & (-1) \left( \frac{1}{n-\Omega} \right) \left( \frac{1}{3r\rho} \right)
\nonumber \\ & \times & 
\frac{d}{dr} \left( r^2 \rho \frac{g}{N^2} \frac{\kappa_\star F_\star}{c_p T} e^{-\kappa_\star p/g} \right)
\nonumber \\ & \times &
Y_{22}(\theta,0) \cos \left[ 2(\phi-\Phi) \right].
\label{eq:xiheq2}
\ee
The quantity in parenthesis and its derivatives are shown in figure
\ref{fig:func}.  It varies mainly as as $p \exp(-\kappa_\star p/g)$,
which has a sharp maximum near the base of the heating layer at
$p=g/\kappa_\star$.  Above the base, this derivative is negative and
$\xi_\phi(\phi) \propto \cos[2(\phi-\Phi)]$ describes motion away from
regions of high entropy, and toward regions of low entropy.  Were there
no return flow, mass would accumulate at $\phi-\Phi=\pi/4,-3\pi/4$, {\it
leading the star}.  For mass ``bumps" leading the star, the quadrupole
moment would act to decrease the planet's spin, pushing it further
from synchronous spin. However, there is a return flow, since below the
base the direction of $\xi_\phi$ reverses (see sign of $df/dr$ in figure
\ref{fig:func}).  As a result, there is no mass accumulation, the density
perturbation is identically zero, and there is no net quadrupole. What is
required to drive asychronous spin is a net flow, integrated over depth,
{\it away} from the the hottest points in the atmosphere.

\subsection{Finite frequency correction}
\label{ss: finite}

In this section we derive an analytic expression for the
density perturbation due to thermal forcing by using perturbation
theory in powers of the frequency. While instructive, this limit is only
applicable when resonances with internal waves are unimportant,
i.e. at forcing periods $\ll 1$ day or $\gg 1$ month.  

Let $\delta p$ and $\xi_r$ denote the complete solution
including both the $\sigma=0$ 
equilibrium tide and the finite frequency corrections.  
To first order in $\sigma^2$, conservation of 
horizontal momentum gives
\be
\delta p & = & -\rho U + \sigma^2 r \rho \xi_h^{\rm (eq)} 
= \delta p^{\rm (eq)} + \sigma^2 r 
\rho \xi_h^{\rm (eq)}.
\label{eq:deltap2}
\ee
Substituting eq.\ref{eq:deltap2} into eq. (\ref{e: mom_r}) allows
us to write the density perturbation as
\be
\delta \rho & \simeq & \delta \rho^{\rm (eq)}
+ \frac{\sigma^2}{g} \left[  \rho \xi_r^{\rm (eq)}
- \frac{d}{dr} \left( \rho r \xi_h^{\rm (eq)} \right) \right].
\ee
For now,  we ignore the tidal potential $U$ and use
eq. (\ref{eq:xiheq}) to write
\be
\delta \rho & = & \frac{\sigma^2}{g} \left[ \rho \xi_r^{\rm (eq)}
- \frac{1}{\ell(\ell+1)} \frac{d^2}{dr^2} \left(  r^2 \rho\xi_r^{\rm (eq)} 
\right) \right].
\label{eq:deltarhocorr}
\ee
This equation shows that finite fluid inertia effects, scaling as
$\sigma^2$, give rise to a nonzero density perturbation
and consequently, a quadrupole moment.

Attempts to directly integrate eq. (\ref{eq:deltarhocorr}) lead to
numerical difficulties due to the second derivative term, which
oscillates with depth. To the extent that $r^{2+\ell}/g$ is constant,
this expression gives a perfect derivative, and the integral would
depend only on the endpoints at the center and surface of the planet,
where the integrand is negligible. In Appendix \ref{a: parts}, we show 
that integration by parts and use of the equations of motion can be
used the obtain a more monatonic integrand with significantly less
cancellation error.

The expression in eq. (\ref{eq:Qparts}) can be estimated analytically in the low frequency limit. If we ignore the 
$\delta \rho \propto \sigma^2$ term, substitute eq. (\ref{eq:xireq}) for
$\xi_r$ and treat the interior mass $m(r) \simeq M_p$
as a constant, we find
\be
Q & = &
\left( \frac{(4+\ell)(3+\ell)}{\ell(\ell+1)} - 1 \right)
\int_0^R dr \rho r^{2+\ell} \frac{\sigma^2}{N^2} \rho_s \frac{\Delta s}{c_p}.
\label{eq:Qanalytic}
\ee
This formula directly gives the quadrupole moment in terms of the applied 
entropy perturbation. Note that the $\ell$-dependent prefactor is equal to 4
for $\ell=2$, and vanishes as $\ell \rightarrow \infty$.

By comparison, \citet{1969Icar...11..356G} assumed constant pressure
and ignored vertical motion to
obtain a density perturbation $\delta \rho/\rho = - \delta T/T=
- \Delta s/c_p$ (for $\rho_s=-1$).\footnote{It is unclear if
\citet{1969Icar...11..356G} assume $\delta p=0$ or $\Delta p=0$.} 
This gives the quadrupole moment
\be
Q^{\rm (GS)} & = &
\int_0^R dr \rho r^{2+\ell} \rho_s \frac{\Delta s}{c_p}.
\label{eq:GSQ}
\ee
The integrand in eq. (\ref{eq:Qanalytic}) has the same
sign as that of eq.\ref{eq:GSQ}, implying a torque which generates asynchronous
spin, but suffers a reduction relative to their expression by a factor 
\be
\left( \frac{(4+\ell)(3+\ell)}{\ell(\ell+1)} - 1 \right)
\frac{\sigma^2}{N^2}.
\label{eq:factor}
\ee

As we will see, our numerical solutions to the boundary value problem
asymptote to the expression in eq. (\ref{eq:Qanalytic}) at long forcing
periods.


\subsection{ numerical results }
\label{sec:numerical}

In this section we present numerical results for the solution of
eq.(\ref{e: mom_r}), (\ref{eq:continuity2}) and (\ref{e: EOS}) for $\delta
p$, $\delta \rho$ and $\xi_r$. To solve these equations, we finite
difference in radius with second order accuracy, and write the resulting
inhomogeneous equation in matrix form as $Mx=B$, where $M$, $x$ and $B$
denote the differential operators, solution vector ($\delta p$, $\delta
\rho$ and $\xi_r$ as a function of radius), and forcing vector (involving
$U$ and $\Delta s$). We find the solution of this linear system using a
band-diagonal solver. The quadrupole moment is then evaluated using 
eq. (\ref{eq:Qparts}).

\begin{figure}
\epsscale{1.2}
\plotone{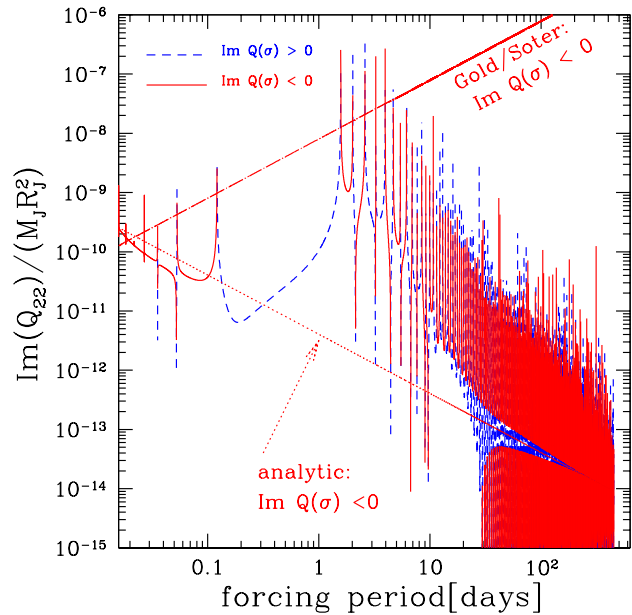}
\caption{Imaginary part of the $\ell=m=2$ quadrupole moment as a
function of forcing period $2\pi/\sigma=\pi/(n-\Omega)$. The quantity
$|Im(Q_{22}(\sigma))|$ is plotted, and short-dashed blue (solid red) lines
show positive (negative) values. For comparison, the approximation
of \citet{1969Icar...11..356G} is shown as the red dot-long dash line, 
while the approximation in eq. (\ref{eq:Qanalytic}) is shown as the red dotted
line. The peaks are due to resonances with normal modes of oscillations,
with g-modes at long period and f-p modes at short period. The base of
the heating layer is at pressure $g/\kappa_\star=1\ {\rm bar}$, and the
base of the radiative zone is at $p_b=100\ {\rm bar}$. The solid red
lines imply a torque pushing the planet away from synchronous spin,
and visa versa for the short-dashed blue lines. }
\label{fig:imq22}
\end{figure}

Figure \ref{fig:imq22} shows the quadrupole moment $Q_{22}(\sigma)$ as
a function of forcing frequency $\sigma$ for the standard boundary
condition in eq. (\ref{eq:bc2}). Forcing is solely by $\Delta s$,
i.e. $U=0$.  The parameters used are $M_p=0.7M_J$, $R_p=1.3R_J$,
$a=0.05\ {\rm AU}$, a solar-type star, base of the heating layer at
$g/\kappa_\star=1\ {\rm bar}$ and base of the radiative zone at
$p_b=100\ {\rm bar}$.  The absolute value of the imaginary part is
plotted. The sign is shown by the line type, solid red line for
$Im(Q_{22}(\sigma))<0$, and dashed blue for $Im(Q_{22}(\sigma))>0$.
The positive forcing frequencies used imply sub-synchronous rotation.
Torque has the same sign as $Im(Q_{22})$, so that $Im(Q_{22})<0$
drives the planet further from the synchronous state, and vice versa.
It is immediately apparent that the response to thermal forcing is
dominated by the low order g-modes at periods $2\pi/\sigma \simeq
1-30\ {\rm days}$. This is in contrast to the gravitational tide,
which has the largest response for the f-mode at $2\pi/\sigma \simeq
0.1\ {\rm days}$.  Consequently, {\it the equilibrium tide limit does
not apply until the frequency is well below the periods of low order
g-modes}. In other words, the low frequency limit of \S \ref{ss: finite}
is not a good approximation until periods $2\pi/\sigma >
30\ {\rm days}$ for the model shown, rather than $2\pi/\sigma \sim
0.1\ {\rm days}$, as one would suspect for the gravitational tide.

For comparison, figure \ref{fig:imq22} shows the GS approximation 
in eq.\ref{eq:GSQ} and the finite frequency correction to the
equilibrium tide given by eq. (\ref{eq:Qanalytic}). The amplitude at
the resonances is set by the frequency spacing used to make the plot;
for finer frequency spacing the peaks would be larger. The solution
asymptotes to eq. (\ref{eq:Qanalytic}) for forcing periods much longer
than the low order g-modes. In the range 1-30 days, the quadrupole
moment is much larger than the value given by
eq. (\ref{eq:Qanalytic}), even midway between resonances. The torque
alternates sign across some resonances, but not others. For short
forcing periods $\sim 0.1-1$ day, the torque acts to synchronize the
spin, reinforcing the dissipative gravitational tide torque. However,
in between the two lowest order g-modes, at periods $\simeq 1-2$ days,
the thermal tide torque is large and acts to torque the planet away
from a synchronous spin state.

\begin{figure}
\epsscale{1.2}
\plotone{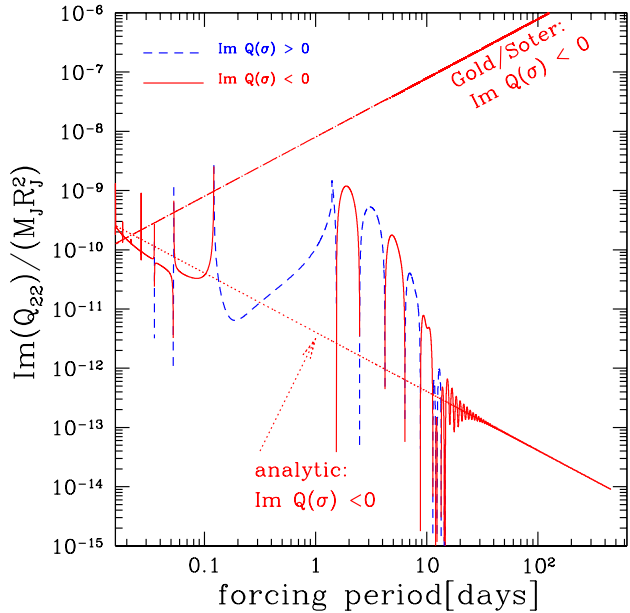}
\caption{Same as figure \ref{fig:imq22} but with an outgoing wave boundary
condition at the upper boundary.}
\label{fig:imq22leak}
\end{figure}

The sharp peaks in figure \ref{fig:imq22} are due to resonant
response. In the absence of damping, the energy of a standing wave
diverges (in linear theory) as the resonance is approached. This is
due to the fact that waves can reflect back and forth in the resonant
cavity, continually being pumped by the forcing. A second limit one
can imagine is that waves with sufficiently short wavelengths at the
top of the atmosphere (above the photosphere) propagate upward,
carrying away their energy.  Figure \ref{fig:imq22leak} shows the
results for the same parameters as in figure \ref{fig:imq22}, the only
modification being the upper boundary condition. Again, there is no
forcing by the gravitational tide ($U=0$). The outgoing wave boundary
condition has eliminated the sharp resonances for high order g-modes. The
exact numerical response agrees well with the analytic formula in
eq. (\ref{eq:Qanalytic}) beyond a forcing period of 1 month. For the
important period range 1 day-1 month, the quadrupole moment is again
1-3 orders of magnitude larger than expected from the analytic
formula, due to the response of low order g-modes -- {\it even for the
leaky upper boundary}.  As we show in the next section, this enhanced
response is due to the good overlap of gravity waves with the
thermal forcing.


\section{Overlap of oscillation modes with
thermal and gravitational forcing}
\label{s: resonant}

At forcing periods $> 1\ {\rm month}$, the frequency dependent
quadrupole moment asymptotically approaches the analytic result in
eq. (\ref{eq:Qanalytic}), derived by treating fluid inertia as a small
correction to the equilibrium tide solution.  However, the range of
forcing periods of interest in determining asynchronous rotation of hot
Jupiters is likely $\la {\rm 1\ month}$.  In this
range of forcing periods, low radial order gravity waves propagate in
the radiative envelope and even the off-resonant response of these waves
increases the quadrupole moment by 1-3 orders of magnitude compared to
eq. (\ref{eq:Qanalytic}).  In what follows, we show that the forced
entropy fluctuations may be understood as an external force that is
exerted upon the planetary fluid. Given the expression for this force,
the standard machinery developed for the purpose of understanding dynamic
gravitational tides (Cowling 1941; Zahn 1970, 1975; Press \& Teukolsky
1977; Goldreich \& Nicholson 1989) can be applied to the thermal forcing
case in order to understand the corresponding quadrupole moments.

\subsection{Forced Oscillator Formalism} \label{ss: formalism}

The fluid fluctuations in the planet can be thought
of as a forced oscillator that obeys
\be
\ddot{\vec{\xi}}+{\bf C}\cdot\vec{\xi}=-{\bf D} \cdot \dot{\vec{\xi}} + \vec{a}
\label{eq:opform}
\ee
where ${\bf C}$ is a self-adjoint operator, responsible
for the adiabatic restoring forces
(Lynden-Bell \& Ostriker 1967) and $\vec{a}=
\vec{a}^G+\vec{a}^S$  
is the sum of the gravitational tidal force
\be
\vec{a}^G=-\grad U
\ee
and the force due to time-dependent insolation, $\vec{a}^S$.
We have also added an ad hoc damping term ${\bf D} \cdot \dot{\vec{\xi}}$.
To compute the form of $\vec{a}^S$, the pressure and
buoyancy forces in eq. (\ref{e: mom_r}) must be
expressed in terms of $\vec{\xi}$ using
eq. \ref{eq:continuity} and \ref{e: EOS} using
\be
\delta p & =& \rho c^2 \left( \frac{\delta \rho}{\rho} - 
\frac{N^2}{g} \xi_r - \rho_s \frac{\Delta s}{c_p}
\right)
\nonumber \\ & = & 
- c^2 \left( \grad \cdot \left( \rho \vec{\xi} \right) + 
\rho \frac{N^2}{g} \xi_r + \rho \rho_s \frac{\Delta s}{c_p}
\right).
\label{eq:deltap}
\ee
Terms involving $\vec{\xi}$ combine to give ${\bf C} \cdot \vec{\xi}$.
The total pressure force contains a non-adiabatic term that results
from time-dependent insolation, which we identify as
\be
\vec{a}^S & = & \frac{1}{\rho} \grad \left( \rho c^2 \rho_s \frac{\Delta s}{c_p} \right)
\ee
The entropy fluctuation $\Delta s$ is specified by eq. 
(\ref{e: first}).

The 
externally forced fluid displacement $\vec{\xi}$ can be 
decomposed into the free adiabatic eigenmodes
$\vec{\xi}_{\alpha}$ as
\be
\vec{\xi}({\bf x},t)=\sum_{\alpha}q_{\alpha}(t)\vec{\xi}_{\alpha}
({\bf x}).
\ee
Here, $q_{\alpha}(t)$ is the time-dependent eigenmode amplitude.
The free adiabatic eigenmodes $\vec{\xi}_\alpha(\vec{x}) e^{-i\sigma_\alpha t}$
with eigenfrequency $\sigma_\alpha$ obey 
\be
-\sigma^2_{\alpha}\,\vec{\xi}_{\alpha}+{\bf C}\cdot\vec{\xi}_{\alpha}
=0.
\ee
The Hermitian nature of the operator ${\bf C}$ implies that the eigenfunctions obey
the orthogonality condition $\int d^3x \rho \vec{\xi}^*_\alpha \cdot \vec{\xi}_\beta
= A_{\alpha \alpha}\delta_{\alpha \beta}$, where $\delta_{\alpha \beta}$ is the delta function
and $A_{\alpha \alpha} = \int d^3x \rho |\vec{\xi}_\alpha|^2$.

By projecting eq. (\ref{eq:opform}) 
onto mode $\alpha$ using the orthogonality relation,
the amplitude $q_\alpha(t)$ obeys the following forced oscillator equation
\be
\ddot{q}_{\alpha}+\sigma^2_{\alpha}q_{\alpha}= - \gamma_\alpha \dot{q}_\alpha 
+  \frac{1}{A_{\alpha \alpha}}
\int d^3 x \rho \vec{\xi}^*_\alpha \cdot \vec{a}(\vec{x},t),
\ee
where $\gamma_\alpha$ is the damping rate. In general, damping can be due to microphysical process
such as viscosity and thermal diffusion. 
It may also represent the loss of energy at the upper boundary due to the outgoing wave boundary condition, which can also then lead to
a broadening of the resonant response.
Plugging in a particular harmonic $\vec{a}(\vec{x},t)=\vec{a}(\vec{x},\sigma)e^{-i\sigma t}$,
the amplitude of the forced response is given by 
\be
q_\alpha(\sigma) & = & \frac{ \int d^3 x \rho \vec{\xi}^*_\alpha \cdot \vec{a}(\vec{x},\sigma) }
{A_{\alpha \alpha} (\sigma_\alpha^2 - \sigma^2-i\sigma\gamma_\alpha) }.
\label{eq:q}
\ee
Note that the mode amplitude $q_{\alpha}(\sigma)$
can be forced by both tidal gravity, which leads to an in phase response,
and the time-dependent insolation, which leads to
an out of phase response.

\subsection{Quadrupole Moment in Terms of Overlap Integrals}

The frequency dependent quadrupole moment is found by combining
the mode amplitude in eq.\ref{eq:q} and the quadrupole moment
from eq.\ref{eq:Qlmdef} to find
\be
&& Q_{\ell m}(\sigma)  \nonumber \\
& = & \sum_\alpha \frac{ \left[ \int d^3 x \rho \vec{\xi}^*_\alpha \cdot
\vec{a}(\vec{x},\sigma) \right]
\left[ \int d^3 x \rho \vec{\xi}_\alpha \cdot \grad \left( r^\ell Y_{\ell m}(\theta,\phi) \right) \right] }
{A_{\alpha \alpha} (\sigma_\alpha^2 - \sigma^2 - i \gamma_\alpha \sigma) }
\label{eq:Qexpansion0}
\\
& = & 
\sum_\alpha Q_{\alpha \ell m}(\sigma) 
\left( \frac{2\sigma_\alpha^2}{\sigma_\alpha^2 - \sigma^2 - i \gamma_\alpha \sigma} \right)
\label{eq:Qexpansion}
\ee 
where we identify the quadrupole moment of mode $\alpha$ as 
\be Q_{\alpha \ell m}(\sigma) & = & \frac{ \left[ \int d^3 x
\rho \vec{\xi}^*_\alpha \cdot \vec{a}(\vec{x},\sigma) \right] \left[
\int d^3 x \rho \vec{\xi}_\alpha \cdot \grad \left( r^\ell Y_{\ell
m}(\theta,\phi) \right) \right] } {2A_{\alpha \alpha} \sigma_\alpha^2
}.
\label{eq:Qalm}
\ee
For forcing frequencies $\sigma \ll \sigma_\alpha$,
eq.\ref{eq:Qexpansion} shows that $2Q_{\alpha \ell m}$
can be identified with the quadrupole moment contribution 
from mode $\alpha$ at low frequency.
Note that $Q_{\alpha \ell m}$
is a physical quantity, since the normalization factors for the 
eigenmodes cancel out.
Also note that the $Q_{\alpha \ell m}$ have frequency dependence solely 
due to $\Delta s \propto \sigma^{-1}$.
This dependence is factored out by writing
\be
{\rm Im}[ Q_{\alpha \ell m}(\sigma) ] & = &
{\rm Im}[ Q_{\alpha \ell m}(\sigma_\alpha) ] 
\left( \frac{\sigma_\alpha}{\sigma} \right).
\label{eq:factorQ}
\ee
The imaginary part of the quadrupole moment, which determines the secular
torque on the orbit and spin, can be written out as
\be
&& {\rm Im}[Q_{\ell m}(\sigma)]  \nonumber \\ 
& = &   \sum_\alpha
\frac{2\sigma_\alpha^2}{(\sigma_\alpha^2-\sigma^2)^2 + \gamma_\alpha^2 \sigma^2}
\nonumber \\ & \times &
\left( {\rm Re}[Q_{\alpha \ell m}(\sigma)] \gamma_\alpha \sigma_\alpha
+ {\rm Im}[Q_{\alpha \ell m}(\sigma)] (\sigma_\alpha^2 - \sigma^2) \right).
\label{eq:ImQmode}
\ee
In the absence of dissipation ($\gamma_\alpha=0$), the gravitational tide
exerts no torque since the response is in phase, and ${\rm Im}[Q_{\alpha
\ell m}(\sigma)]=0$.  Including dissipation ($\gamma_\alpha \neq 0$),
the gravitational tide gives rise to a torque due to the first term
in eq. (\ref{eq:ImQmode}). By contrast, the thermal tide response
is inherently out of phase.  Hence even when $\gamma_\alpha=0$, the
thermal tide causes a secular torque. Inclusion of damping for the
thermal tide will act to prevent divergent response at resonances,
due to the denominator of the Lorentzian in eq. (\ref{eq:ImQmode}).

\begin{figure}
\epsscale{1.2}
\plotone{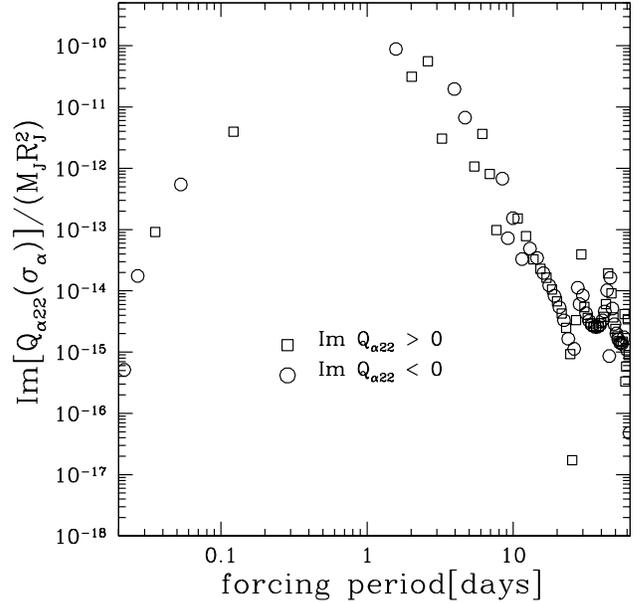}
\caption{ 
Quadrupole moments $Q_{\alpha 22}$ for the oscillation
modes. G-modes have periods longer than 1 day, and f-p modes have
periods shorter than 1 day.  The plot shows
$Q_{\alpha 22}$ for $\ell=m=2$ for each oscillation mode for the sum
in eq. (\ref{eq:Qexpansion}) for the frequency dependent
quadrupole. The base of the radiative zone is at $p_b=100 {\rm bar}$,
and the base of the heated layer is at $g/\kappa_\star=1 {\rm
bar}$. Other parameters are the same as in figure the same as in
figure \ref{fig:imq22}. The circles (squares) show modes with
$Im(Q_{\alpha 2 2})<0$ ($Im(Q_{\alpha 2 2})>0$).  A radial resolution
of 9447 grid points was used to compute the results shown in this
plot. The results are numerically converged for most of the modes
shown even at much lower resolution.  }
\label{fig:imq22mode}
\end{figure}

\begin{figure}
\epsscale{1.2}
\plotone{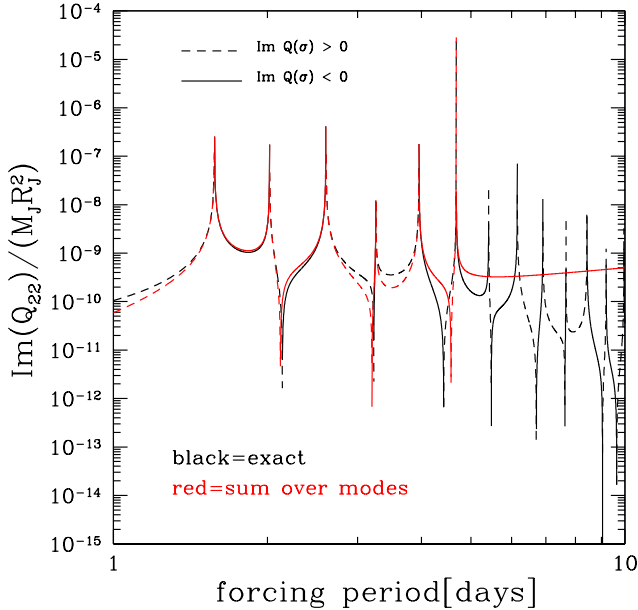} 
\caption{ 
Comparison of the numerical response (black lines) versus the sum over
modes in eq.  \ref{eq:Qexpansion} and \ref{eq:factorQ} (red
lines). The coefficients $Q_{\alpha 22}(\sigma_\alpha)$ for each mode
are shown in figure \ref{fig:imq22mode}, and the response is for the
case shown in figure \ref{fig:imq22}. Only modes in the period range
1-5 days were used to make the red lines; in particular, no f or p
modes were used. Note the sum over modes agrees well for low order
g-modes, even midway between resonances, implying the off-resonant
response of the f and p-modes is weak compared to that of low order
g-modes. Agreement could be improved by including additional modes in
the sum.  }
\label{fig:sum}
\end{figure}

Figure \ref{fig:imq22mode} shows the quadrupole moments $Q_{\alpha
22}(\sigma_\alpha)$ for each mode as defined in eq. (\ref{eq:Qexpansion}) and (\ref{eq:Qalm})
for the same parameters used to find the frequency dependent
quadrupole moment in figure \ref{fig:imq22}.
These quadrupole moments
were computed by fitting the $Q_{22}(\sigma)$ near resonance using
eq. (\ref{eq:Qexpansion}) in order to read off the coefficient $Q_{\alpha
22}(\sigma_\alpha)$. It is immediately apparent that the dominant overlap
is with the lowest order g-mode at period $\simeq 1\ {\rm day}$. The f
and p mode quadrupoles are down by 1-2 orders of magnitude.

To verify the equivalence between the solution of the boundary value
problem and the sum over eigenmodes, the frequency dependent response
$Q_{22}(\sigma)$ is compared to the sum over eigenmodes from
eq. (\ref{eq:Qexpansion}) in figure \ref{fig:sum}. Damping is neglected, i.e.
the standard boundary condition in eq. (\ref{eq:bc2}) is used. Despite the fact that
only a handful of low order g-modes are used in the sum, and that the
f and p modes are not included, there is good agreement with the
solution to the full boundary value problem.

The difference resulting from the two different upper boundary conditions as
shown in figures \ref{fig:imq22} and \ref{fig:imq22leak}
can be understood using eq.\ref{eq:ImQmode}. Figure \ref{fig:imq22}
has $\gamma_\alpha=0$, and hence divergences appear in the quadrupole
moment near the frequencies of internal standing waves. In figure
\ref{fig:imq22leak}, damping is introduced through the outgoing wave
boundary condition. In fact this damping is so large that $\gamma_\alpha \sim \sigma_\alpha$,
and the Lorentzian factors $\sigma_\alpha^4/( (\sigma_\alpha^2-\sigma^2)^2 + \gamma_\alpha^2 \sigma^2)
= {\rm O}(1)$. For Lorenztian factor of order unity, the induced quadrupole moment is
of order the overlap integral $Q_{\alpha \ell m}(\sigma_\alpha)$.

Even for large damping rates $\gamma_\alpha
\sim \sigma_\alpha$, the low order g-modes still dominate the response
since their overlap integrals with the thermal tide force are far larger
than the f and p-modes (see figure \ref{fig:imq22mode}). In the duration of
one wave travel time in the radiative layer, the wave can build up
enough energy to dominate the response. The overlap integrals in figure
\ref{fig:imq22mode} are larger than near-equilibrium tide result from
eq. (\ref{eq:Qanalytic}) in the period range 1 day - 1 month, precisely the 
region where enhanced response is seen in figure  \ref{fig:imq22leak}.

The spatial profiles of thermal and gravitational forcing differ
greatly from one another. Thermal forcing is localized in physical
space to the outer layers of the planet, resembling a delta function
in space. By contrast, the gravitational tide most closely resembles
the f-mode, and can be thought of as a delta function in momentum space
\citep{1994ApJ...432..296R}.  In other words, since thermal forcing is
narrow in physical space, the fluid response, in terms of normal modes,
is broad in momentum space, and visa versa for the gravitational tide.
However, the $g-$modes which propagate within the stably stratified
envelope dominate since their frequencies are comparable to forcing
frequencies of interest, and their eigenfunctions are confined to the
radiative layer where the thermal forcing occurs. The frequencies of f
and p-modes are too high to be resonant, and their eigenfunctions extend
over the entire planet.


\section{ on the neglect of the Coriolis force, thermal diffusion, and
background flows }
\label{sec:approx}

In an attempt to elucidate the origin of quadrupole moments from thermal
forcing, we have made a number of simplifying assumptions so as not to
obscure the basic physics. In this section we comment on how our results
might be changed, both qualitatively and quantitatively, by relaxing
these assumptions.

\subsection{the Coriolis force}

Consider the effect of uniform rotation, with spin frequency $\Omega$.
When $\sigma \la \Omega$, the Coriolis force becomes important for the
fluid response and we anticipate several additional effects due its
presence. First, the gravity wave dispersion relation and fluid
motions are altered.  For $\sigma \sim \Omega$, we expect the change
in the quadrupole to be at the factor of a few level in the ``bump"
between 1-30 days.  Second, new wave families arise, mainly restored
by the Coriolis force -- the Rossby waves, or inertial waves. These
waves can propagate throughout both the radiative envelope and the
convective core. We expect that, for sufficiently rapid
supersynchronous rotation, Rossby resonances may give rise to a larger
quadrupole moment.  In the perturbation theory calculation of \S
\ref{ss: finite}, we expect {\it larger} quadrupole moment when the
spin is nearly synchronous, since the Coriolis terms $ \propto \sigma
\Omega$ will be larger than the inertia terms $\propto \sigma^2$. In
figures \ref{fig:imq22} and \ref{fig:imq22leak} this will lead to the
``analytic" line becoming flat, instead of decreasing to longer
forcing period.

For sufficiently large quadrupole moment, the equilibrium spin will be
asychronous enough that $\sigma \ga \Omega$, and we expect our results
to be approximately quantitatively correct.  Substitution of the
quadrupole moment in figure \ref{fig:imq22leak} into
eq. \ref{eq:quad_compar}, for ${\cal Q}_p=10^6$ we find $\sigma\sim n$
in the period range of $\sim$ a few days, implying the spin frequency,
orbital frequency and forcing frequency are all comparable in torque
equilibrium. In this event, the Coriolis force is important, but will
only change the frequencies and eigenfunctions of the gravity waves at
the order unity level.


\subsection{thermal diffusion}

The heating function in eq.\ref{eq:epsilon} describes the law 
of exponential attenuation of stellar radiation. By setting this
to occur at a pressure $\sim 1\ {\rm bar}$, we are assuming there
are no absorbers high in the atmosphere. Until recently, this 
assumption was standard, but there may now be evidence of
an inversion layer due to such absorbers in a fraction of 
observed hot Jupiters \citep{2008ApJ...678.1419F}. In that
case, incorporating thermal forcing by the boundary condition 
used in \citet{2009MNRAS.395..422G}
may be more appropriate. \footnote{ If one considers a
two-frequency, two-angle approximation to the radiative transfer,
both the ``greenhouse" and the ``stratosphere" cases may
be included.}

The entropy perturbation in eq.\ref{e: first} assumes that the 
forcing period is much shorter than the thermal time in the heated
layer, so that radiation diffusion may be ignored. In fact,
thermal diffusion and damping of downward propagating waves is
expected, as in the solutions of \citet{2009MNRAS.395..422G}.
Furthermore, a shorter diffusion time may partially ``erase"
temperature perturbations, reducing the buoyancy force upon which
gravity waves rely. This may cause wave damping or reflection 
near the optical or infrared photospheres. By choosing the ``perfect
reflector" and ``leaky" upper boundary conditions in section 
\ref{sec:numerical} we attempted to simulate two extreme limits
of the effects of radiative diffusion. While thermal diffusion may
decrease the amplitude of gravity waves near the photosphere, deeper
in the envelope the thermal time becomes sufficiently long so that
damping can be ignored.

For forcing period of order a few days, the thermal time at the base
of the heating layer is of order the forcing period. Therefore, the
approximation that the layer possesses a large thermal inertia is only
approximately valid. Another effect we have ignored is that nonlinear
effects become important when the forcing period is longer than the thermal time.

\subsection{fluid flow in the background state}

We now comment on the possible role of fluid motion such as zonal
flows in the background state. Fluid motion in the
background state does not preclude the existence of the gravity (or
inertial) wave response as described in this paper.  Ignoring shearing
of the velocity field, such a flow would simulate uniform rotation, so
that the Coriolis force becomes important. However, when there is
velocity shear in the background state, we expect the wave motions to
be significantly modified when the shear rate is comparable to the
wave frequency. Since inertial wave response relies on angular
momentum gradients, we expect these waves to persist in the presence
of velocity shear.  Most atmospheric circulation simulations for the
hot Jupiters (e.g. \citealt{2008ApJ...673..513D},
\citealt{2009ApJ...699..564S}) assume synchronous rotation, and find
powerful advective flows. Relatively few simulations with asychronous
rotation have been performed, but the recently results by
\citet{2009ApJ...699..564S} are useful in this context. They found
that when the degree of asynchronous spin was increased, the flow
velocities were decreased. It remains to be seen what winds are
induced for rotation rates appropriate for the torque equilibrium
described in section \ref{s:motivation}.  All current studies of
atmospheric circulation in hot Jupiters ignore the thermal tide
torques.  From the analysis presented in this work, it seems likely
that the qualitative and quantitative outcome of these simulations will
change once thermal tidal torques are included.



\section{Summary}

\label{s: summary}

We considered the ability of a planet, subjected to time-dependent
stellar irradiation, to develop net quadrupole moments. The existence of such 
quadrupole moments allows the stellar tidal field to exert torques on the 
planet, possibly creating asynchronous spin. Such asynchronous spin could lead
to large tidal heating rates, through the dissipative gravitational tide,
perhaps sufficient to power the large observed radii.

In \S \ref{ss: zero_freq} and \ref{ss: finite}, the fluid equations
are solved treating inertia as a small parameter. At zeroth order in
the forcing frequency, a circulation pattern is found from hot to cold
at small depths $p \la g/\kappa_\star$, and a return flow from cold to hot
at larger depth $p \ga g/\kappa_\star$. There is no density
perturbation associated with this flow and therefore, zero quadrupole
moment. The inclusion of finite inertia as a small perturbation gives
rise to a
finite density perturbation and therefore, quadrupole moment.
In the assumed limit of small fluid
inertia, the phase of the quadrupole is found to have the correct sign
to induce asynchronous rotation in the planet, similar to the
work of \citet{1969Icar...11..356G}.  However, the magnitude of this
quadrupole is reduced by a factor $\sim 4(\sigma/N)^2$. A propagating
wave-like response of any form is eliminated in this ``near-equilibrium
tide'' approximation.  The value of the resulting quadrupole moment is
insensitive to boundary conditions and depends solely on the local
forcing in the atmosphere. Numerical calculations presented in figure
\ref{fig:imq22} and \ref{fig:imq22leak} confirm that the quadrupole
moment approaches this analytic limit at long forcing periods.

In \S \ref{sec:numerical}, the fluid equations are solved as a boundary
value problem. Two different upper boundary conditions are used, which
effectively treat the radiative envelope as a perfect resonant cavity, or
as a cavity with an open upper lid, allowing waves to be radiated upward.
Using both boundary conditions, which may bracket the true result
including radiative transfer effects, it is found that the quadrupole
moments in the period range 1 day - 1 month are far larger than the
non-resonant near-equilibrium tide calculation in eq. \ref{eq:Qanalytic}
by 1-3 orders of magnitude. The reason for this enhanced response
is discussed in \S \ref{s: resonant}. Unlike the gravitational tidal
forcing, time-dependent thermal forcing and low order envelope g-modes
couple well due a favorable spatial overlap.

In section \ref{s:motivation}, we argued that the magnitude of the
thermal tide torques using the formula of \citet{1969Icar...11..356G}
is sufficient to generate large deviations from synchronous rotation
as well as large tidal heating rates, easily sufficient to power the observed
planet radii. Our numerical results in \S \ref{sec:numerical} show that
the full numerical solutions for the torque can indeed have the correct 
sign to generate asynchronous spin for some ranges of forcing (rotation)
frequency. The magnitude of the torques is smaller than the
\citet{1969Icar...11..356G} result by a factor of a few,
but still large enough to be interesting. We expect that the inclusion of the 
Coriolis force will {\it increase} the torque at long forcing periods,
since the Coriolis force is larger than fluid inertia in that limit.

In future investigations, we plan on including the effects of the
Coriolis force and thermal diffusion in the calculation of the
quadrupole moments.  We expect that both physical effects could lead
to qualitative and quantitative changes to the non-resonant and resonant
thermal tidal response.  The resulting torques will then be used to
solve for the equilibrium values of the planet's spin, radius and
eccentricity.




\acknowledgements We thank Peter Goldreich for helpful discussions.
Also, we thank Jeremy Goodman, Gordan Ogilvie and Pin-Gao Gu for
raising the issue of isostatic adjustment as well as Yanqin Wu for
discussions of the upper boundary condition. We also thank the referee
for comments that improved the paper.


\appendix

\section{Accurate evaluation of the quadrupole moment: integration by parts}
\label{a: parts}

As discussed in section \ref{ss: finite}, direct integration of
$r^\ell \delta \rho$ to find the quadrupole moment is subject to large
cancellation errors. This behavior is somewhat surprising, and does not occur
for the gravitational tide quadrupole moment. The origin of the problem is
that $\delta \rho$ is proportional to the second derivative of a certain
quantity, resulting in large cancellations during the integration. We
noticed this cancellation error during numerical resolution studies in
which the results did not converge rapidly with increasing resolution.

Here, we show that this cancellation error can be ameliorated
through successive integration by parts, and use of the equations of
motion. By using the exact equations of motion, this method does not
introduce approximations.  Integration by parts leads to an integrand
that is much more monatonic, leading to better convergence.


The quadrupole moment is given by
\be
Q &= & \int_0^R dr r^{2+\ell} \delta \rho.
\ee
By substituting for $\delta \rho $ from eq. (\ref{e: mom_r}) gives
\be
Q &= & 
\int_0^R dr r^{2+\ell} \frac{\rho}{g} \left( \sigma^2 \xi_r - \frac{1}{\rho} \frac{d\delta p}{dr}
- \frac{dU}{dr} \right).
\ee
Integrating the pressure gradient term by parts, discarding the (small) boundary terms, 
and solving eq.\ref{eq:continuity2} for $\delta p$, we find
\be
Q &= & \int_0^R dr r^{2+\ell} \frac{\rho}{g} \left( \sigma^2 \xi_r
- \frac{dU}{dr} \right)
\nonumber  + 
 \frac{\sigma^2}{\ell(\ell+1)} \int_0^R dr \frac{d}{dr} \left( \frac{r^{2+\ell}}{g} \right)
\left[ r^2 \delta \rho + \frac{d}{dr} \left( r^2 \rho \xi_r \right)
\nonumber \right.  -  \left.
 \frac{\ell(\ell+1)}{\sigma^2} \rho U
\right].
\ee
Integrating the $d/dr(r^2 \rho \xi_r)$ term by parts and simplifying gives
the final result
\be
Q & = & \int_0^R dr r^{2+\ell} \delta \rho^{\rm (eq)}
+ 
\sigma^2 \int_0^R dr \left[ \rho \xi_r \left\{ \frac{r^{2+\ell}}{g}
- \frac{r^2}{\ell(\ell+1)} \frac{d^2}{dr^2} \left( \frac{r^{2+\ell}}{g} \right) \right\}
\right.  +  \left.
 \frac{r^2}{\ell(\ell+1)}  \frac{d}{dr} \left( \frac{r^{2+\ell}}{g} \right) \delta \rho
\right].
\label{eq:Qparts}
\ee
The first term in eq. (\ref{eq:Qparts}) is just the equilibrium
$\sigma^2=0$ quadrupole from gravitational forcing
alone -- see eq. (\ref{e: rho_eq}).  Although
the $\delta \rho$ term in eq. (\ref{eq:Qparts}) may be subject to cancellation
error, this term should be smaller by a factor $\sigma^2 r/g$ compared to
the other terms after integration. Since the equilibrium tide component of $\xi_r$ is monatonic, 
the terms involving $\xi_r$ do not suffer cancellation error.  We find numerically that integrand
is far more monotonic than $r^{2+\ell} \delta \rho$, and is less subject to
cancellation error. Throughout, we use eq. (\ref{eq:Qparts}) when
numerically evaluating the quadrupole moment that results from thermal forcing.



\begin{thebibliography}{}
\bibitem[Arras \& Bildsten(2006)]{2006ApJ...650..394A} Arras, P., \& Bildsten, L.\ 2006, \apj, 650, 394 
\bibitem[Burrows et al.(2000)]{2000ApJ...534L..97B} Burrows, A., Guillot, T., Hubbard, W.~B., Marley, M.~S., Saumon, D., Lunine, J.~I., \& Sudarsky, D.\ 2000, \apjl, 534, L97 
\bibitem[Cowling(1941)]{1941MNRAS.101..367C} Cowling, T.~G.\ 1941, \mnras, 101, 367 
\bibitem[Dobbs-Dixon \& Lin(2008)]{2008ApJ...673..513D} Dobbs-Dixon, I., \& Lin, D.~N.~C.\ 2008, \apj, 673, 513 
\bibitem[Fortney et al.(2008)]{2008ApJ...678.1419F} Fortney, J.~J., Lodders, K., Marley, M.~S., \& Freedman, R.~S.\ 2008, \apj, 678, 1419 
\bibitem[Gold \& Soter(1969)]{1969Icar...11..356G} Gold, T., \& Soter, S.\ 1969, Icarus, 11, 356 
\bibitem[Goldreich \& Kumar(1990)]{1990ApJ...363..694G} Goldreich, P., \& Kumar, P.\ 1990, \apj, 363, 694 
\bibitem[Goldreich \& Nicholson(1989)]{1989ApJ...342.1079G} Goldreich, P., \& Nicholson, P.~D.\ 1989, \apj, 342, 1079 
\bibitem[Goldreich \& Soter(1966)]{1966Icar....5..375G} Goldreich, P., \& Soter, S.\ 1966, Icarus, 5, 375 
\bibitem[Goodman(2009)]{2009arXiv0901.3279G} Goodman, J.\ 2009, arXiv:0901.3279 
\bibitem[Gu \& Ogilvie(2009)]{2009MNRAS.395..422G} Gu, P.-G., \& Ogilvie, G.~I.\ 2009, \mnras, 395, 422 
\bibitem[Newcomb (1962)]{newcomb62} Newcomb, W.~A. 1962, Nuclear Fusion: Supplement Part 2, Vienna: International Atomic Energy Agency, 451.
\bibitem[Press \& Teukolsky(1977)]{1977ApJ...213..183P} Press, W.~H., \& Teukolsky, S.~A.\ 1977, \apj, 213, 183 
\bibitem[Reisenegger(1994)]{1994ApJ...432..296R} Reisenegger, A.\ 1994, \apj, 432, 296 
\bibitem[Showman et al.(2009)]{2009ApJ...699..564S} Showman, A.~P., Fortney, J.~J., Lian, Y., Marley, M.~S., Freedman, R.~S., Knutson, H.~A., \& Charbonneau, D.\ 2009, \apj, 699, 564 
\bibitem[Unno et al.(1989)]{1989nos..book.....U} Unno, W., Osaki, Y., Ando, H., Saio, H., 
\& Shibahashi, H.\ 1989, Nonradial oscillations of stars, Tokyo: University of Tokyo Press, 1989, 2nd ed.,  
\bibitem[Zahn(1970)]{1970A&A.....4..452Z} Zahn, J.~P.\ 1970, \aap, 4, 452 
\bibitem[Zahn(1975)]{1975A&A....41..329Z} Zahn, J.-P.\ 1975, \aap, 41, 329 
\end{thebibliography}
\end{document}